\documentclass[%
 reprint,
superscriptaddress,
 amsmath,amssymb,
 aps,
]{revtex4-2}
\usepackage{comment}
\usepackage{mathtools}
\usepackage{graphicx}
\usepackage{dcolumn}
\usepackage{float}
\usepackage{soul}
\usepackage{cancel}
\usepackage{bm}

\usepackage[colorlinks]{hyperref} 
\begin{document}

\preprint{APS/123-QED}

\title{Sensory adaptation in a continuum model of bacterial chemotaxis - working range, cost-accuracy relation, and coupled systems}


\author{Vansh Kharbanda}
\author{Benedikt Sabass}%
\affiliation{Faculty of Physics and Center for NanoScience, Ludwig-Maximilians-Universität München}
\affiliation{Institute for Infectious Diseases and Zoonoses, Department of Veterinary Sciences, \\Ludwig-Maximilians-Universität München}




\date{\today}

\begin{abstract}
Sensory adaptation enables organisms to adjust their perception in a changing environment. A paradigm is bacterial chemotaxis, where the output activity of chemoreceptors is adapted to different baseline concentrations via receptor methylation. The range of internal receptor states limits the stimulus magnitude to which these systems can adapt. Here, we employ a highly idealized, Langevin-equation based model to study how the finite range of state variables affects the adaptation accuracy and the energy dissipation in individual and coupled systems. Maintaining an adaptive state requires constant energy dissipation. We show that the steady-state dissipation rate increases approximately linearly with the adaptation accuracy for varying stimulus magnitudes in the so-called perfect adaptation limit. This result complements the well-known logarithmic cost-accuracy relationship for varying chemical driving. Next, we study linearly coupled pairs of sensory units. We find that the interaction reduces the dissipation rate per unit and affects the overall cost-accuracy relationship. A coupling of the slow methylation variables results in a better accuracy than a coupling of activities. Overall, the findings highlight the significance of both the working range and collective operation mode as crucial design factors that impact the accuracy and energy expenditure of molecular adaptation networks.

\end{abstract}

\maketitle


\section{Introduction}
The surroundings of all living organisms constantly change and diverse mechanical, electrical, and chemical signaling pathways enable organisms to continuously detect these changes~\citep{Laughlin1989,nakatani1991light,berg2004coli,Hendrik2004}. Reliable detection of changes in different signal backgrounds requires the adjustment of the sensitivity, which is called sensory adaptation~\cite{alon}. A paradigm of sensory adaptation is chemotaxis of the bacterium \textit{Escherichia coli} (\textit{E. coli})~\cite{JAdler}. Chemotaxis allows \textit{E. coli} to navigate up or down chemical concentration gradients, e.g., to find an environment with higher sugar concentrations. Chemotactic movement is a result of a stochastic control of the rotation direction of the flagella that drive bacterial swimming~\cite{berg2004coli}. \textit{E. coli} swims in a pattern that comprises of successive straight runs and tumble phases, during which it reorients. As the bacterium moves towards higher concentrations of a chemoattractant, the tumbling frequency is reduced, which leads to increasingly long runs that take the cell in the direction of higher concentration. Conversely, the tumbling frequency is increased as the bacterium swims away from a repellent.\\
The presence of ligands is sensed by transmembrane receptors that are called methyl-accepting chemotaxis proteins. Depending on ligand binding, receptors can either be in an active or an inactive state. Chemoattractant binding inhibits receptor activity, which suppresses tumbling by lowering the phosphorylation state of a response regulator CheY~\cite{ultrasensitive}. Simultaneously, an adaptation of the receptors to the ligand concentration occurs via methylation or demethylation of their four to five glutamate residues with the help of the methyltransferase CheR and the methylesterase CheB, respectively~\cite{Adler79,Knox_1986}. CheR preferentially methylates inactive receptors and increases receptor activity through methylation if receptor activity is reduced by an attractant. CheB preferentially demethylates active receptors and thereby lowers activity upon removal of attractants. Consequently, adaptation restores the receptor activities to their original value after an initial response to a ligand concentration change~\cite{alon}. This feedback mechanism can in principle ensure a precise and robust detection of concentration gradients, regardless of the ambient background concentration of ligands. However, experiments also revealed a limited working range of adaptation with imprecise adaptation to very high concentrations of attractants~\cite{berg1972chemotaxis,meir2010precision,min2012chemotactic}. This limitation most likely results from a saturation of available methylation sites~\cite{neumann2014imprecision}. { While the impact of a limited working range on the sensory performance have been studied theoretically
\citep{meir2010precision,neumann2014imprecision}, it is less well-understood how the energetic cost of adaptation depends on the working range of the mechanisms. Chemosensory adaptation that is realized by a finite number of receptor methylation sites can be idealized as a dynamical process with a bounded range of internal variables. How such a bounded internal phase space affects the energetics of nonlinear biological processes is generally an open question. This issue motivates our study of sensory adaptation with a highly idealized model where the interplay of phase-space boundaries of internal variables, energetics, and adaptive performance can be studied.}\\
Chemotaxis receptor molecules form clusters at the cell membrane~\cite{gegner1992assembly, kim2002dynamic}, where individual molecules form homodimers and three such homodimers form a trimer~\cite{briegel2012bacterial, liu2012molecular, haselwandter2014role}. A receptor core complex, the minimal {unit} of signaling, contains two trimers and shows a non-linear dose-response~\cite{Parkinson_2016}. Assemblies of multiple chemoreceptors work in a cooperative manner to enhance the sensitivity of the response to subtle concentration gradients~\cite{interactions_between_receptors,Sourjik_2004, Wingreen2011}. 
{On a molecular level, cooperativity} can result from enzymes that change the methylation state of the dimer they are bound to as well as the methylation state of neighboring dimers, e.g., CheR and CheB~\cite{le1997methylation, li2005adaptational}. 
 {The molecular mechanisms of receptor cooperativity, as well as its functional advantages, e.g., to achieve optimal signal-to-noise ratios, have been studied extensively. However, the impact of receptor cooperativity on the energy consumption required for adaptation is hardly explored and warrants further theoretical research.}

A variety of models for sensory adaptation in chemotaxis have been proposed over the past decades~\citep{BL_model,alon1999robustness,yi2000robust,NetworkTopology, Wingreen_2006,Endres_2006, Tu_review}. {Idealized feedback systems can be represented by only a few coupled differential equations~\cite{tu2008modeling,Shimizu2}. However, chemosensory adaptation is usually modeled as a stochastic process since the} number of molecules constituting a sensory apparatus in individual cells is typically quite small and thermal noise and active processes result in noisy signals. This noise affects the reliability of information transmission between an input signal and the output~\citep{bialek2005physical, MutualInfoBWInpandOut, hartich2015nonequilibrium} and thereby determines the precision of the cellular response. {However, fluctuations are also intimately connected with the energetic cost of adaptation.} From an information-theoretic viewpoint, sensory adaptation always entails a thermodynamic cost since it implies an irreversible erasure of stored information along with a measurement~\cite{Sartori_2014}. Moreover, the two fundamental models of noisy cellular adaptation, namely the negative feedback mechanism in chemotaxis and the incoherent type-1 feed-forward mechanism, {only show adaptation if the systems' states are held in non-equilibrium ~\citep{Lan2012,CostofResponse}. The energetic cost of maintaining a non-equilibrium state required for adaptation can be related to sensory performance by an energy-speed-accuracy relation}~\cite{Lan2012}, which states that the dissipation rate is proportional to the negative logarithm of the adaptation error, both measured in steady state after application of a constant stimulus. 


Most work on the energy-speed-accuracy relationship is based on a discrete description of receptor states using master equations~\cite{Lan2012}, {for which the} original relationship can be generalized~\citep{Wang_2015}. {Here, one assumes two discrete states for the fast receptor activity and four to five discrete states that represent the slowly changing methylation level that acts as a a negative control element in \textit{E. coli} chemotaxis.} Alternatively, the receptor states can be represented in a continuum picture using Langevin equations for both the response and the control variables. Based on simulation results for such a model, the generality of the energy-speed-accuracy relation has been questioned~\citep{Maes2017}. The Langevin-equation based model is highly idealized and not as directly applicable to chemotaxis as the discrete model. However, it has the advantage that the adaptation process and the associated fluxes of probability can be clearly visualized in a two-dimensional phase space. This facilitates an intuitive understanding of the adaptation process.
{In principle, one may even study a hybrid model where the states of the control variable of the receptor are discrete and the response variable is continuous, although the technical advantage of such an approach is not obvious.}

In this article, we take advantage of the clear representation offered by the Langevin-equation approach to study the relationships between adaptation, working range, and energetic cost for individual and coupled systems. Unlike earlier studies~\cite{Lan2012,Wang_2015}, we directly simulate the Langevin dynamics, taking into account the phase-space boundaries. Plotting the adaptation error against the dissipation rate for varying input signals, we find that the finite working range results in a linear cost-accuracy relation, {which highlights the importance of the finite working range of the internal variables on the performance and the cost of adaptation.} This result complements the well-known logarithmic energy-speed-accuracy relation~\cite{Lan2012}. {Also, to underpin our numerical findings, we derive an analytical expression for the energetic cost of adaptation as a function of input stimulus strength}. Furthermore, motivated by the discovery that receptor clusters consist of minimal sensory units with a non-linear response~\cite{Parkinson_2016}, we consider pairs of receptors in the spirit of Ref.~\cite{ENDRES200933} and ask how a linear coupling of the sensor states affects the overall cost-accuracy relationship. {To this end, we study two examples of coupled systems and find that, in general, a linear coupling reduces the cost of adaptation without affecting its accuracy.}

\section{A Single Sensory System}
\subsection{Model Introduction}
An idealized model of bacterial adaptation can be built from a three-node network with a negative feedback. We focus on a Langevin-type model suggested in Ref.~\cite{Lan2012}, where all variables are assumed to have non-negative, real values.
The model comprises of a variable $a$ for the rapidly changing receptor activity  and a slowly changing control variable $m$, representing the receptor methylation state. The activity $a$ responds to changes in an input stimulus $s$ and also represents the output of the system. 

\begin{figure*}
    \centering
    \includegraphics[width = \linewidth]{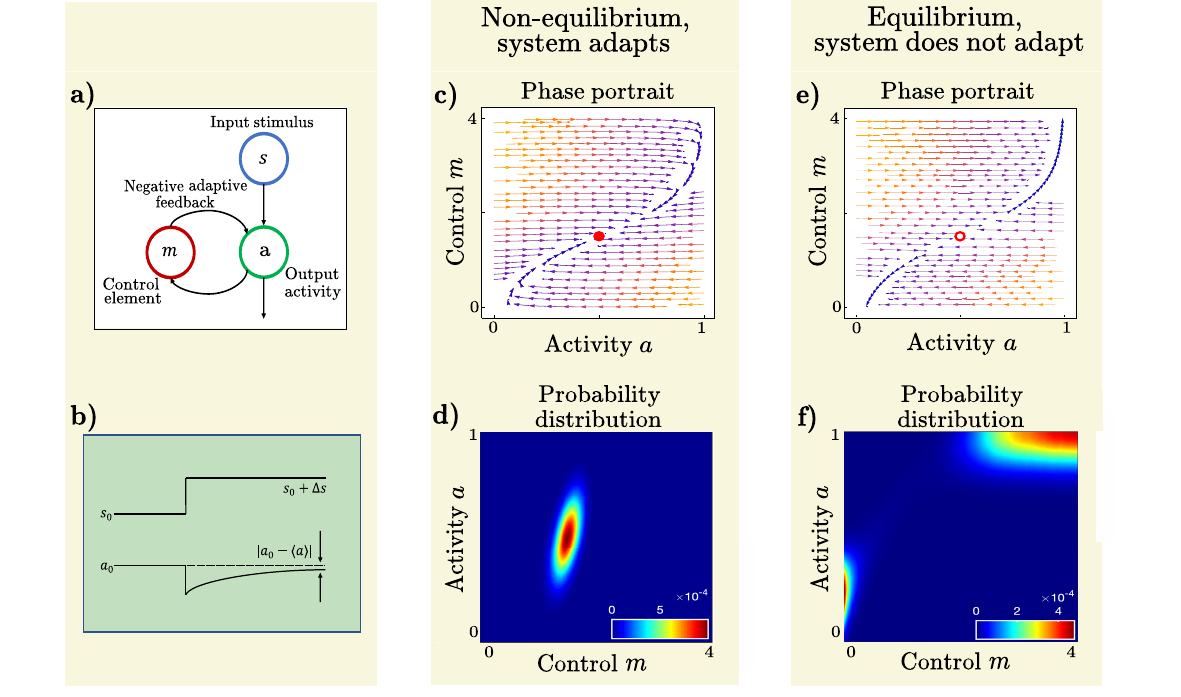}
    \caption{\textbf{(a)}~A minimal three-node network with a negative feedback loop. \textbf{(b)}~Following a change of the stimulus $s$, the instantaneous response of the activity $a$ is cancelled by a gradual change in the control variable $m$. The adaptation error is defined as the deviation of the expectation value of $a$ from its setpoint $a_0$ in the stationary limit. \textbf{(c)}~Phase portrait of the dynamics in the absence of noise. For $\beta = 1$, the fixed point of the deterministic dynamics $(a_0,m^*)$ is stable and represents the adaptive state of the system. \textbf{(d)}~Steady-state probability distribution of the system variables in the presence of noise. The deterministic fixed point lies at the center of the probability distribution, that is, $(\langle a \rangle_{\text{ss}}, \langle m \rangle_{\text{ss}}) = (a_0,m^*)$. \textbf{(e)}~For $\beta = 0$, the chemical driving vanishes and the fixed point $(a_0,m^*)$ is an unstable saddle node. \textbf{(f)}~In the presence of noise, the state of the non-adaptive system is driven to the corners of the phase space. \textit{Simulation parameters: $s = 20$, $\omega_a = 50, \omega_m = 5, T = 0.01.$ For (f), we chose $s = 3$ to illustrate that the state localizes in both corners of the phase space.}}
    \label{fig:panel1}
\end{figure*}

Following a stimulus, a change in the activity $a$ drives a slow compensatory change in the control variable $m$, which ultimately brings $a$ back to the stimulus-independent setpoint $a_0$, see FIGs.~\ref{fig:panel1}a and \ref{fig:panel1}b. The dynamics of the activity and the control (methylation) variables is described by a pair of coupled Langevin equations as
\begin{equation}
\dot{a} = F_{a}(a,m,s) + \eta_a(t); \quad \dot{m} = F_{m}(a,m,s) + \eta_m(t).
\label{langevinsystem}
\end{equation}
\noindent Here, $\eta_a(t)$ and $\eta_m(t)$ both represent white noise. Denoting expectation values as $\langle\ldots\rangle$, the noise satisfies $\langle \eta_i(t)\eta_j(t') \rangle = 2 \Delta_{i}\delta_{ij}\delta(t'-t)$ where the indices $i$ and $j$ represent $a$ or $m$. The functions $F_a$ and $F_m$ provide a lumped description of the nonlinear biochemical reactions that govern the dynamics of fast response and slow adaptation, respectively \cite{Lan2012}.

The Fokker-Planck equation (FPE) that corresponds to the Langevin equations~(\ref{langevinsystem}) describes the evolution of the phase-space probability distribution function (PDF) $P \equiv P(a,m,t)$ as
\begin{equation}
\partial_t P = -\partial_a J_a - \partial_m J_m ,
\label{FPE}
\end{equation}
where $J_a \equiv F_aP - \Delta_a\partial_a P$ and $J_m \equiv F_mP - \Delta_m\partial_m P$ are the components of the probability current $\mathbf{J}$. 

{It can be shown that the system must obey the following condition to satisfy detailed-balance}
\begin{equation}
\Delta_m \partial_m F_a = \Delta_a \partial_a F_m,
\label{DBC}
\end{equation}
{which is also derived in the Supplementary Material.}
This detailed-balance condition for the negative feedback control mechanism implies that adaptation cannot be realized in equilibrium and is inherently dissipative~\cite{AdapatationinLivingSystems}.

We specify the functional form of the forces driving the dynamics of $a$ and $m$ as
\begin{equation}
F_a(a,m,s) = -\omega_a[a - G(s,m)]
\label{Force_a}
\end{equation}
and
\begin{equation}
F_m(a,m,s) = -\omega_m(a - a_0)[\beta - (1 - \beta)C\partial_{m}G(s,m)],
\label{Force_m}
\end{equation}
where
\begin{equation}
    C \equiv \frac{\Delta_m/\omega_m}{\Delta_a/\omega_a}
\end{equation}
is assumed to be a constant and $\beta \in [0,1]$ is a parameter that determines whether the system is in equilibrium, $\beta = 0$, or out-of-equilibrium, $\beta > 0$. The rate of the response and adaptation processes are determined in Eqs.~(\ref{Force_a}, \ref{Force_m}) by the constants $\omega_a$ and $\omega_m$, respectively. We assume that
\begin{equation}
\omega_a \gg \omega_m,
\label{time-scale-sep}
\end{equation}
such that the time scales of response and adaptation are well-separated.

The function $G(s,m)$ is chosen to be a Hill equation in the Michaelis-Menten limit, i.e., with Hill coefficient $H = 1$ as
\begin{equation}
G(s,m) = (1+s/(K_0e^{2m})^H)^{-1},
\label{Gsm}
\end{equation}
where we set $K_0=1$. The functional form chosen in Eq.~(\ref{Gsm}) ensures that $\partial_m G > 0$ and $\partial_s G < 0$, which is required for the negative feedback mechanism \cite{Lan2012}.\\
{In a deterministic setting, for} 
any $\beta$, there exists a fixed point where the two nullclines of the deterministic system intersect. 
Consequently, there is a value of the control variable $m=m^*$ that satisfies
\begin{equation}
G(s,m^*) = a_0.
\label{mstar_eqn}
\end{equation}
Moreover, there exists a $\beta_c$ such that the fixed point is stable only if $\beta > \beta_c$, where $0 < \beta_c < 1$ \cite{Lan2012}. Thus, stable adaptation is only achieved if $\beta > \beta_c$. Phase portraits of the system with stable fixed points for $\beta = 1$ and unstable fixed points for $\beta = 0$ are shown in FIGs.~\ref{fig:panel1}c and \ref{fig:panel1}e, respectively. The corresponding probability distributions are presented in FIGs.~\ref{fig:panel1}d and \ref{fig:panel1}f. 
Throughout this article, we focus for simplicity on the case of $\beta = 1$, which is called the fully adaptative state which represents the perfect adaptation limit.
\subsection{Assumptions \& Definitions}
\subsubsection{Phase space \& Adaptation Error}
For the model described in Eqs.~(\ref{langevinsystem}), the activity $a$ and control state $m$ are stochastic variables that lie in a phase space denoted as $\bar{\Omega} \equiv [0,1]\times[0,m_0]$, with $m_0$ corresponding, e.g., to the maximum number of methylation sites in a receptor. To solve the FPE, Eq.~(\ref{FPE}), on the interior of the phase space, $\Omega~\equiv~\bar{\Omega} \setminus \partial\Omega$, a set of well-defined boundary conditions is required. For this model, we impose reflective boundary conditions, which state that the current vector obeys
\begin{equation}
   \mathbf{n}\cdot\mathbf{J} = 0 \quad \text{for}\,\,\, (a,m)^T\in\partial\Omega.
   \label{BC}
\end{equation}
where $\mathbf{n}$ is the outward-pointing vector normal to the boundary $\partial \Omega$.

The setpoint for the activity $a_0$ is assumed to be a constant. Since the function $F_m$ is designed to be proportional to $(a-a_0)$, the generalized force driving methylation changes vanishes at the setpoint. The integral feedback mechanism ensures that the statistical average of the activity 
\begin{equation}
\langle a(t) \rangle = \iint_{\Omega} a\,P(a,m,t)\,\text{d}a\,\text{d}m
\label{mean_a}
\end{equation}
approaches $a_0$ in the steady state~\cite{Lan2012}. Denoting the steady-state average as $\langle \ldots \rangle_{\mathrm{ss}}$, the adaptation error 
\begin{equation}
\epsilon \equiv |1 - \langle a \rangle_{\mathrm{ss}}/a_0|
\label{adaptationerror}
\end{equation}
quantifies the accuracy of adaptation in the stationary limit after the decay of transient responses to a signal.

\subsubsection{Effective Temperature \& Energy Dissipation Rate}
\noindent
In equilibrium, the fluctuation-dissipation theorem (FDT) can be used to relate the temperature of the system to the noise amplitude and the damping in the system~\cite{Kampen2007}. In the following, we set the Boltzmann constant to unity, $k_{\text{B}} \equiv 1$, and define the effective temperatures pertaining to the stochastic variables of the system in Eq.~(\ref{langevinsystem}) as follows
\begin{equation}
T_{\text{eff(a)}} \equiv \frac{\Delta_a}{\omega_a},\quad T_{\text{eff(m)}} \equiv \frac{\Delta_m}{\omega_m}. 
\end{equation}
We assume that the response and adaptation processes described in Eq.~(\ref{langevinsystem}) are coupled to the same heat bath. Therefore, the effective temperatures are equal $T_{\text{eff(m)}}=T_{\text{eff(a)}}=T$, which entails that $C =1$.

Since the product of temperature and the entropy production rate equals the dissipation rate in a stationary, isothermal setting, see Ref.~\cite{Seifert2012}, the stationary dissipation rate can be expressed as
\begin{equation}
\dot{W}_{\text{diss}} = T \iint_{\Omega} \bigg[\frac{J_a^2}{\Delta_a P} + \frac{J_m^2}{\Delta_m P} \bigg]\, \text{d}a\,\text{d}m.
\label{integralDissip}
\end{equation}
This equation for the dissipation rate is only valid for a system with single heat bath and temperature $T$. For a system with multiple heat baths, additional terms must be included in Eq.~(\ref{integralDissip}) to account for the heat flux between reservoirs~\cite{Sekimoto2010, Harada_2005}.

An analytical approximation for the dissipation rate was given in Ref.~\cite{Lan2012}, where Eq.~(\ref{time-scale-sep}) and Laplace's integral method were used to estimate the integrals in Eq.~(\ref{integralDissip}), leading to the following expression
\begin{equation}
\begin{split}
\dot{W}_{\text{diss}}\approx\dot{W}^{\text{a}}_{\text{diss}} & = \frac{\omega_m T}{C} \big( 1 + C\,\partial_m G(s,m)\big)^2\big|_{m=m^*} \\
    &= \omega_m T (1 +  \partial_m G(s,m))^2\big|_{m=m^*},
    \label{W^a}
\end{split}
\end{equation}
where $C=1$, the superscript ``a'' stands for the analytical approximation proposed in Ref.~\cite{Lan2012}, and $m^*$ is the $m$-coordinate of the fixed point of the deterministic system i.e., it satisfies Eq.~(\ref{mstar_eqn}).

{Also, employing the framework of stochastic thermodynamics~\cite{Sekimoto2010, Seifert2012}, the dissipation rate for an individual trajectory ``t'' of the system can be expressed as}
\begin{equation}
   \dot{W}_{\text{t}} = T\big(\Delta_a^{-1}F_a\circ\dot{a} + \Delta_m^{-1}F_m\circ\dot{m} \big),
   \label{DissipationTrajectory}
\end{equation}
where $\circ$ indicates a product in the Stratonovich sense~\cite{Sekimoto2010}. For an ensemble of trajectories, we take the statistical average of Eq.~(\ref{DissipationTrajectory}) to obtain an expression for the dissipation rate in stationary state as
\begin{equation}
   \dot{W}_{\text{diss}} = \big\langle T\big(\Delta_a^{-1}F_a\circ\dot{a} + \Delta_m^{-1}F_m\circ\dot{m} \big)\big\rangle_{\mathrm{ss}}.
   \label{DissipationStochastic}
\end{equation}
In contrast to Eq.~(\ref{integralDissip}), Eq.~(\ref{DissipationStochastic}) can be straight-forwardly evaluated using the trajectories generated by simulations of the stochastic dynamics.
\subsection{Simulation Methods}
To simulate the system  of Langevin equations~(\ref{langevinsystem}), we employ an explicit first-order method as described in Ref.~\cite{kloeden2011numerical}, {where the boundary conditions can be accounted for by adding a term that ensures normal reflections at the boundary \cite{skorokhod}. }However, to make the effect of the boundaries explicit in analytical calculations of the dissipation rate, we mostly employ smooth potentials that approximate the no-flux boundary conditions of the system. The potentials are 

\begin{equation}
    V(a,m) \equiv \frac{1}{\omega_a}V_b(a,a_{\text{max}}) + \frac{1}{\omega_m}V_b(m,m_{\text{max}}),
\end{equation}
where $(a,m) \in \Omega$ and the function $V_b$ with arguments $a$ or $m$ is given by
\begin{equation}
    V_b(r,r_{\text{max}}) = V_0\big[\exp(\frac{-r}{\sigma}) + \exp(\frac{r-r_{\text{max}}}{\sigma}) \big].
\end{equation}
\noindent The potential $V_b$ is a superposition of exponential functions which approximate hard, reflecting walls in the limit of small $\sigma>0$. {Non-zero values of $\sigma$ ensure that the potential is smooth and bounded for $r\in(0,r_{\text{max}}).$} To test if the finite range of the wall potentials affects the  simulation results, we also perform simulations using ``reflection operators", {the details of which can be found in the Supplementary Material.} {The results of computations using the reflection operators} are also presented in the Supplementary Material (FIG. S2) and agree with {results that were obtained from computations that make use of the boundary potentials.}

\subsection{Analytical and Numerical Results}
In this section, we study how the finite variable-range in a fully adaptive system leads to a non-zero adaptation error that increases with the magnitude of the stimulus. Concomitantly, the boundaries reduce dissipation, which results in a novel functional form of the cost-accuracy trade-off.

\subsubsection{Simulation Results: Dependence of Adaptation Error on System Variables}
Since the system consists of fluctuating state variables, understanding its adaptive performance requires a consideration of the probability distribution of the variables. The probability distribution is {affected by the presence of phase-space boundaries that determine the range of the system variables}. We define the bulk of the phase space, denoted by $\mathcal{B} \subseteq \Omega \subset \bar{\Omega}$, as the largest subset of the phase space in which the {boundary-induced changes of }the probability distribution {do} not result in measurable changes in adaptation error.

The capacity of a system to adapt after a change in stimulus $s$ is limited by the phase-space boundaries that determine the maximum and minimum values of system variables~\cite{Lan2017, MelloTu2003}. In the present model, this claim can be proven by separating the adaptation error, Eq.~(\ref{adaptationerror}), into three contributions,

\begin{equation}
    \epsilon = |\epsilon_1 - \epsilon_2 + \epsilon_3|
    \label{epsilon_split},
\end{equation}
with
\small
\begin{subequations}
    \begin{align}
        \epsilon_1 &=  \frac{1}{\beta \omega_m a_0}\int_0^1\mathrm{d}a\;\Delta_m P(a,0)\mathrm{d}a \label{eq5.26a}, \\
        \epsilon_2 &= \frac{1}{\beta \omega_m a_0}\int_0^1\mathrm{d}a\;\Delta_m P(a,m_0)\mathrm{d}a \label{eq5.26b},\\
        \epsilon_3 &= \frac{1-\beta}{\beta a_0}\iint_{\Omega} \mathrm{d}a\,\mathrm{d}m \; P(a,m)\partial_mG(s,m)(a-a_0), \label{eq5.26c}
    \end{align}
    \label{split_epsilon}%
\end{subequations}
\normalsize
which are obtained using the time-scale separation of adaptation and response. In Eq.~(\ref{epsilon_split}), $\epsilon_1$ and $\epsilon_2$ are contributions due to the phase-space boundaries, while $\epsilon_3$ accounts for the error in the bulk. For perfect adaptation, $\beta = 1$, the bulk contribution vanishes~\cite{Lan2012}. Therefore, in a fully adaptive state, any error in the stationary value of the adapted state results from the limited range of the internal variables.

\begin{figure}
    \centering
    \includegraphics[width = \linewidth]{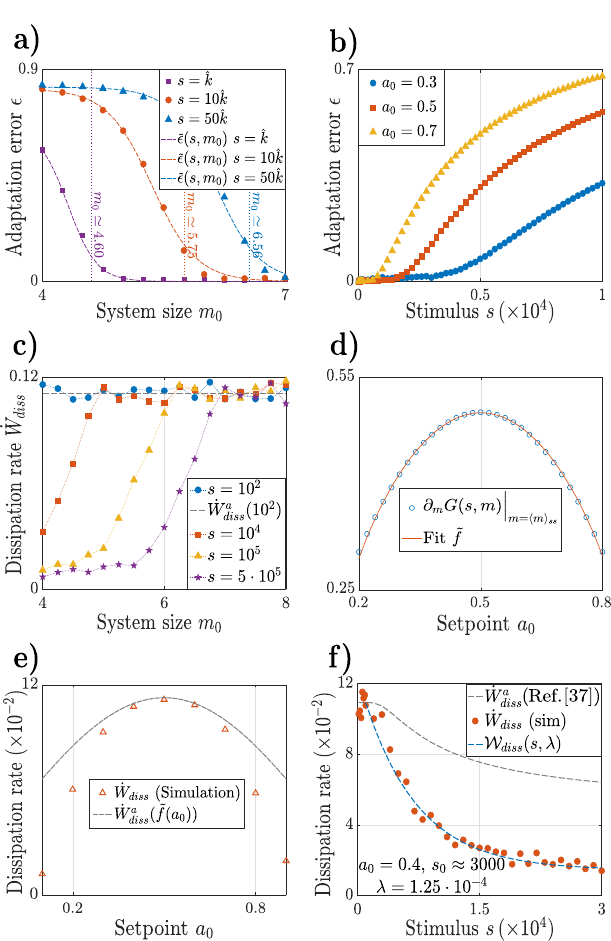}
    \caption{\textbf{(a)}~Dependence of the adaptation error on the range of the control (methylation) variable $m \in [0,m_0]$, with $\hat{k} = 10^4$. For large $m_0$, the error vanishes for a fully adaptive system since the state rarely encounters the boundary at $m_0$. Simulation results can be fitted with an exponential form, see Eq.~(\ref{error_m0_fit}). Vertical lines show values of $m_0$ at which the corresponding deterministic systems produce an error of $5\%$. 
    \textbf{(b)}~Adaptation error as a function of stimulus strength at fixed system size with $m_0 = 4$.
    \textbf{(c)}~The dissipation rate $\dot{W}_{\text{diss}}$ of the adaptive system decreases with reduced system size if the state encounters the phase-space boundary. For small stimuli, $s = 10^2$, the state remains in the center of the space and the dissipation is is independent of system size.
    \textbf{(d, e)}~The function $\partial_{m}G(s,m)|_{m=\langle m \rangle_{\text{ss}}}$ and the dissipation rate are roughly quadratic functions of the setpoint distance from the neutral state $a_0=0.5$, which lies in between the active state ($a = 1$) and the inactive state ($a=0$) of the system.
    \textbf{(f)}~Dissipation rate as a function of the stimulus strength $s$. Simulation results $\dot{W}_{\text{diss}}$ are compared with an analytical approximation $\dot{W}^{\text{a}}_{\text{diss}}$ and an empirical formula $\mathcal{W}(s,\lambda)$.
    The three quantities agree for small stimuli $s<s_0$. Otherwise, $\dot{W}^\text{a}_{\text{diss}}$ differs from the simulation results due to the breakdown of the analytical approximation at the boundaries.} 
    \label{fig:Panel3}
\end{figure}

Since in a perfectly adaptive state the adaptation error is only non-zero near the boundaries, the adaptation error for a large stimulus $s$ depends on the maximum value that the methylation variable can assume. This maximum value is determined by $m_0$. Results from numerical computations displayed in FIG.~\ref{fig:Panel3}a show that the dependence of the adaptation error on $m_0$ can be fitted by a Fermi-Dirac-like function as
\begin{equation}
    \tilde{\epsilon} (s,m_0) := \frac{1}{c_1+e^{c_2(m_0 - c_3)}} = \frac{1}{c_1+e^{\tilde{c_2}\hat{k}(m_0 - c_3)/s}},
    \label{error_m0_fit}%
\end{equation}
where $c_1,c_2 =  \tilde{c}_2 \frac{\hat{k}}{s}$,and $c_3$ are fit parameters and $\hat{k} = 10^4$ is a constant scale factor. In FIG.~\ref{fig:Panel3}a, the values of $m_0$ are shown at which the analytical solution of the deterministic system predicts an error of $5\%$ for different $s$. These values indicate the points where the boundary effects start to become significant for a given stimulus. Figure~\ref{fig:Panel3}b illustrates that the adaptation error also depends on the setpoint $a_0$. A larger setpoint results in an increased adaptation error.

\subsubsection{Simulation Results: Dissipation Rate}
If the stimulus magnitude $s$ exceeds a threshold $s_0$, the solution of Eq.~(\ref{mstar_eqn}) yields a fixed point $m^* >m_0$ that lies outside the variable domain, $m^* \not\in \bar{\Omega}$. Then, the steady-state expectation value $\langle m \rangle_{\text{ss}}$ deviates from $m^*$. To compute the dissipation rate $\dot{W}^\text{a}_{\text{diss}}$, see Eq.~(\ref{W^a}), outside the bulk $\mathcal{B}$, we replace $m^*$ with $\langle m \rangle_{\text{ss}}$. A comparison of $m^*$ and $\langle m \rangle_{\text{ss}}$ as a function of the stimulus $s$ can be found in the Supplementary Material, FIG.~S3. 

The interaction of state trajectories with phase-space boundaries reduces the system's energy dissipation rate in the stationary state. As shown in FIG.~\ref{fig:Panel3}c, for a small value of the input stimulus, $s = 10^2$, the dissipation rate of the system remains constant for varying $m_0$. However, for larger $s$ the dissipation of the system starts to decrease when the probability distribution touches the boundary at $m_0$. Furthermore, when $s$ is fixed, changing the setpoint $a_0$ of the activity variable 
also affects the energy cost, as moving the distribution along the $a-$axis introduces effects of the boundaries at $a = 0$ or $ a=1$. The plot in FIG.~\ref{fig:Panel3}e shows that the dissipation rate decreases symmetrically as $a_0$ is moved away from the center at $a_0 = 0.5$.

\subsubsection{Breakdown of Approximation for the Dissipation Rate and Empirical Formula}
Comparison of the dissipation rate in simulations, $\dot{W}_{\text{diss}}$, with the approximate expression $\dot{W}^{\text{a}}_{\text{diss}}$, Eq.~(\ref{W^a}), shows that the latter only holds if the setpoint is close to the center of the phase space, $a_0 = 0.5$, see FIG.~\ref{fig:Panel3}e. Thus, the analytical approximation holds only if the effect of the boundaries is very small whereas $\dot{W}_{\text{diss}} \leq \dot{W}^a_{\text{diss}}$ otherwise.

The behavior of the dissipation rate $\dot{W}_{\text{diss}}$ can be further analyzed by defining a threshold value $s_0$ for the input stimulus. Above this threshold, the probability distribution is sufficiently {altered by} the boundary at $m = m_0$ to change the dissipation rate by more than some fixed percentage. For $a_0\in (0.3,0.7)$, see FIG.~\ref{fig:Panel3}e, and a large enough value of $s$ such that $m^*\not\approx 0$, we conjecture the following expression for the dissipation rate
\begin{equation}
    \mathcal{W}_{\text{diss}} \coloneqq  \begin{cases}
    \dot{W}^{\text{a}}_{\text{diss}}& s < s_0 \\
    K_0e^{-\lambda(s-s_0)} + \epsilon_0(1 - e^{-\lambda(s-s_0)}) & s\geq s_0
    \end{cases}
    \label{eq7.5}
\end{equation}
where $K_0 \equiv \dot{W}^{\text{a}}_{\text{diss}}\big|_{s=s_0}$ is the dissipation coefficient and $\epsilon_0  \equiv \min\limits_{s\in (0,\infty )}\dot{W}_{\text{diss}}$. Results from the empirical expression $\mathcal{W}_{\text{diss}}$ and $\dot{W}_{\text{diss}}$ are shown in FIG.~\ref{fig:Panel3}f. The expression in Eq.~(\ref{eq7.5}) describes the numerical data well,
while the approximation $\dot{W}^a_{\text{diss}}$ fails. 

\subsubsection{Cost-Accuracy Relation}
What is the energetic cost of accurate adaptation? A well-known energy-speed-accuracy relation, proven in Ref.~\cite{Lan2012}, asserts a monotonous increase of the dissipation rate with adaptation accuracy for varying chemical driving force. More precisely, the energy-speed-accuracy relationship states that the dissipation rate is proportional to the negative logarithm of the error. See the Supplementary Material (FIG. S4) for a recapitulation of this result in our simulations where the chemical driving is controlled by the value of $\beta$. Optimal accuracy is reached in the limit of perfect adaptation where the dissipation is maximal.

\begin{figure}
    \centering
    \includegraphics[width = \linewidth]{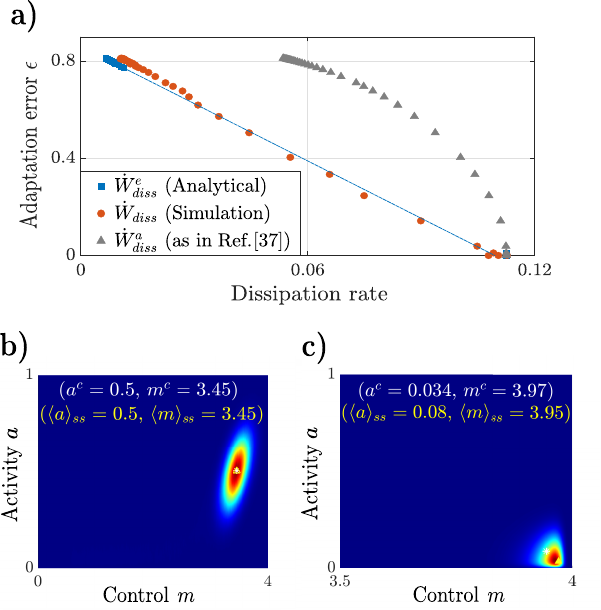}
    \caption{\textbf{(a)}~Cost-accuracy relation for varying stimulus strength. The adaptation error $\epsilon$ is linearly related to the the dissipation rate $\dot{W}_{\text{diss}}$ at different values of $s$. Analytical results $\dot{W}^\text{e}_{\text{diss}}$ estimating the dissipation rate at weak and strong stimuli, $s<s_0$ and $s\gg s_0$, agree with the simulation data. The steepest decent approximation used for $\dot{W}^{\text{a}}_{\text{diss}}$ breaks down for strong stimuli.
    \textbf{(b)}~Steady-state probability distribution after application of a weak stimulus, $s=10^3<s_0$. The distribution resembles a Gaussian and the averages co-localize with the solution of Eq.~(\ref{constraints}), $(a^c,m^c)$. \textbf{(c)}~State probability distribution after application of a strong stimulus, $s=5\cdot 10^4 >s_0$. The distribution is skewed and the averages do not co-localize with the maximum of the distribution $(a^c,m^c)$.
    \textit{Simulation parameters: $V_0 = 1/2,\, \omega_m = 5,\, \omega_a = 50, \sigma = 0.01,\, T = 0.01$.}}
    \label{fig:panel4}
\end{figure}

To understand how the finite range of internal variables affects the relationship between dissipation and accuracy, we vary here the magnitude of the external stimulus $s$. We focus on the limit of a fully adaptive model, i.e., $\beta = 1$. As shown in FIG.~\ref{fig:panel4}a, the adaptation error obtained in simulations decreases linearly with a growing dissipation rate. In other words, the accuracy increases with the energetic cost also if the error solely results from a finite variable range. This relationship is thus qualitatively similar to the energy-speed-accuracy relation, but the functional forms and implications are different.

To buttress our numerical result for the cost-accuracy relation, we next perform analytical calculations of the dissipation rate. As illustrated in FIGs.~\ref{fig:Panel3}e and \ref{fig:Panel3}f, the validity of the approximate expression $\dot{W}^{\text{a}}_{\text{diss}}$ breaks down when significant adaptation errors occur close to phase-space boundaries.  Thus, we derive a new equation for the dissipation rate by explicitly taking into account the boundaries of the variable domain.
\noindent 
We consider the system of Langevin equations with any smooth respulsive boundary potential $V$ as 
\begin{subequations}
\begin{equation}
    \dot{a} = F_a(a,m,s) - \omega_a\partial_a V + \eta_a,
\end{equation} 
\begin{equation}
    \dot{m} = F_m(a,m,s) - \omega_m\partial_m V + \eta_m.
\end{equation} 
\end{subequations}
These equations are linearized around any point near the phase-space boundary $(a^c,m^c)$ of their deterministic counterparts. Using Eqs.~(\ref{Force_a}, \ref{Force_m}) and $\beta = 1$, we obtain
\begin{subequations}
\begin{equation}
\begin{split}
    \dot{a} = &-\omega_a(a - a^c)\big[1+\partial_a^2V\big|_{a=a^c}\big]\\ &+\omega_a \partial_m G(s,m)\big|_{m=m^c}(m-m^c) + \eta_a,
\end{split}
\end{equation}
\begin{equation}
    \dot{m} = -\omega_m(a-a^c) - \omega_m\partial_m^2 V\big|_{m=m^c} (m-m^c) + \eta_m,
\end{equation}
\label{linear_LE}%
\end{subequations}
{where the higher order terms of the expansion are ignored and the leading terms in the Langevin equations drop out, as we define the fixed point} $(a^c,m^c)$ {to satisfy},
\begin{subequations}
\begin{equation}
     \quad G(s,m^c) - \partial_aV\big|_{a=a^c} - a^c =0,
\end{equation}
\begin{equation}
    (a^c-a_0)+\partial_mV\big|_{m=m^c}=0.
\end{equation}
\label{constraints}%
\end{subequations}
The above two equations represent a modified version of Eq.~(\ref{mstar_eqn}). The boundary potentials ensure that the solution $m^c$ exists for all values of the input and that $m^c = m^*$ for $s<s_0$.

The fixed point $(a^c,m^c)$ corresponds to the point of maximum likelihood - the point in the phase space where the probability density function reaches its maximum value, see FIGs.~\ref{fig:panel4}b and~\ref{fig:panel4}c. The {FPE corresponding to the} linearized equations~(\ref{linear_LE}) is solved by a Gaussian ansatz. Substitution of the solution into the expression for the dissipation rate, Eq.~(\ref{integralDissip}), yields
\begin{equation}
    \dot{W}^{\text{e}}_{\text{diss}} = T_{\text{eff}}\frac{(1+\partial_mG(s,m))^2\omega_a\omega_m}{\omega_a + \omega_m\partial_m^2 V+\omega_a\partial_a^2 V}\bigg|_{\substack{a = a^c \\ m = m^c}}.
    \label{approxwork}
\end{equation}
Here, $\partial_a^2 V$ and $\partial_m^2 V$ are both non-negative due to the choice of boundary potentials. The superscript ``e'' in $\dot{W}^{\text{e}}_{\text{diss}}$ indicates that this expression arises from a linear expansion of the governing equations of the system. 

{The approximation in Eq.~(\ref{approxwork}) is only valid when either $s < s_0$ or $s\to \infty$ because the state probability distribution can be approximated by a Gaussian distribution only for such values of $s$, see FIG.~\ref{fig:panel4}b. For other values of $s$, the distribution is skewed and an expansion around the point of maximum likelihood results in larger errors.}

As illustrated in the Supplementary Material, simulations with the chosen exponential boundary potentials yield very similar results to simulations employing a reflection operator with very small timesteps. Thus, the results are expected to also hold for reflective boundaries. Also, a master-equation based chemotaxis receptor model with discrete molecule states yields qualitatively the same cost-accuracy relation for varying stimulus strength as shown for the Langevin-equation based model in this section, see Supplementary Material.
\subsection{Discussion of Results for a Single Sensory System}
Using a minimal Langevin-equation-based model for adaptation~\cite{Lan2012}, we analyze how a limited range of internal variables affects adaptation accuracy and energetic cost. The model consists of a stochastic variable $a$ called activity, a slow, stochastic control variable $m$, and an externally controlled stimulus $s$. In this work, the adaptation error $\epsilon$ is defined as the systematic, relative deviation of the activity $a$ from its setpoint $a_0$, see Eq.~(\ref{adaptationerror}). The adaptation process can be visualized by following the motion of the system state $(a,m)$ in the rectangular variable domain $\bar{\Omega}$ with $a \in [ 0,1 ]$ and $m \in [ 0,m_0 ]$. For the fully adaptive model considered here, the adaptation error solely results from the limited range of the internal variables $a$ and $m$.  To explain the onset of the error $\epsilon$, consider the state to be initially at its setpoint with $\langle a \rangle = a_0$. After a sudden increase of the stimulus $s$, $\langle m \rangle$ 
slowly shifts to higher values to bring $\langle a \rangle$ back to its setpoint. However, if $s$ becomes larger than a certain input-threshold $s_0$, such that the control variable approaches its upper limit $m_0$, the reflective boundaries {alter} the probability distribution, see FIGs.~\ref{fig:panel4}c and \ref{fig:panel1}d. Then the control variable cannot increase sufficiently to counteract the effect of the external stimulus on the activity $a$ and the perfect adaptability of the system is compromised.
Unlike the error due to the limited range of the control variable, which results from the deterministic dynamics of the system, the error resulting from finite domain of $a$ is a consequence of the finite, non-zero variance of the noise in the system. 
Also, as shown in FIG.~\ref{fig:Panel3}a, the adaptation error saturates for very large stimulus $s$ to a maximum value $\lim_{s\to\infty}\epsilon \equiv \epsilon_{\text{max}} < 1$. $\epsilon_{\text{max}}$ is independent of both the setpoint and the system size as long as $m_0$ is finite. The limit on the maximum error results from the skewness of the probability distribution.

The considered energetic cost of adaptation is the steady-state dissipation rate after a change of the input signal. Perpetual dissipation in the adaptive system results from the non-equilibrium forces that maintain the state $(a,m)$ close to a fixed point in the presence of fluctuations. The fluctuations drive the state away from its fixed point and work is required to bring it back. Upon strong stimulation, the probability distribution for $(a,m)$ moves to the vicinity of the boundaries. Intuitively, the effect of boundaries can be interpreted here as passive forces that also maintain the state at a fixed point. More formally, Eq.~(\ref{BC}) ensures that the normal component of the fluxes at the boundary vanishes, thereby reducing the contribution of these fluxes to the integrand in Eq.~(\ref{integralDissip}). Consequently, the dissipation rate decreases at the boundary, see FIG.~\ref{fig:Panel3}f.

A plot of adaptation error against dissipation rate as a function of stimulus strength yields a linear cost-accuracy relationship as a result of the limited range of internal variables. {Thus, our result pertain to the behavior of a sensory system in a regime where its internal variables reach their physical limits. In that case, we see that the system is unable to adapt, but it also reduces the energy it uses for adaptation in a linear fashion.} By contrast, variation of the chemical driving force yields a logarithmic cost-accuracy relationship, called the energy-speed-accuracy relation~\cite{Lan2012}. Thus, the functional form of cost-accuracy relations depends on the nature of the internal constraints imposed on the adaptation process. We emphasize, however, that the energy-speed-accuracy relation can directly imply a design trade-off for cells since the adaptive performance can be improved by investing more energy. This is not the case for the cost-accuracy relationship studied here, which depends on the range of internal variables and the external stimulus. {In order to turn our results into a proper} trade-off relationship, one would need to determine how the cost of enlarging the state space, i.e., to generate molecules with more methylation sites in the case of chemotaxis, improves chemotaxis.

\section{Interacting Systems}
\begin{figure*}
    \centering
    \includegraphics[width = \linewidth]{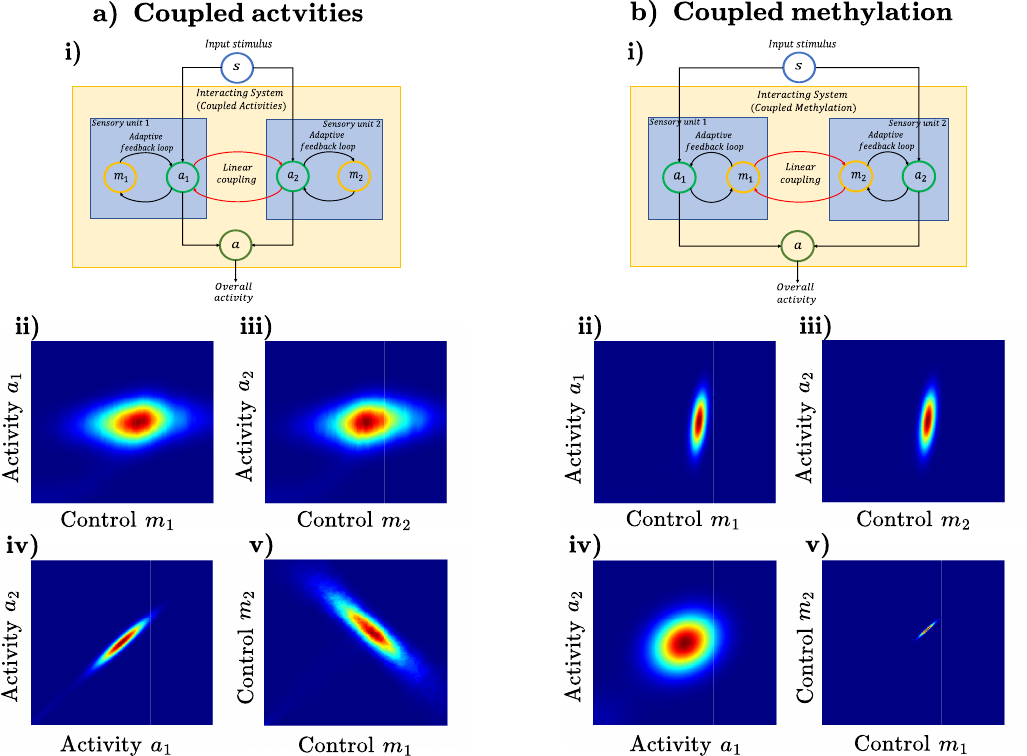}
    \caption{\textbf{(a.(i), b.(i))}~Composite systems consisting of two sensory units, where each realizes an adaptive feedback loop. The units are linearly coupled either via their activity or their control (methylation) variables. The overall activity $a$ of the system is the average of the individual activities $a_1$ and $a_2$. \textbf{(a.(ii) - a(v))}~Steady-state probability distribution $P$ visualized on two-dimensional planes in the phase space for coupled activities. A large coupling constant $\omega_{12}> \omega_{a_{\{1,2\}}}$ leads to a strong correlation of the rapidly changing activities $a_1$ and $a_2$. The control variables $m_1$ and $m_2$ are also strongly correlated. As a result, the distributions on the $a_{i}-m_{i}$ planes for the individual units are also {altered}.
    \textbf{(b.(ii) - b.(v))}~State probability distribution $P$ visualized on two-dimensional planes in the phase space for coupled control variables. Due to the slow dynamics of the control variables in the individual units determined by $\omega_{a_{\{1,2\}}} \gg \omega_{m_{\{1,2\}}}$, the variances of the $m_{\{1,2\}}$ are smaller than the variances of the $a_{\{1,2\}}$. As a result, even if a large coupling constant enforces a strong correlation of $m_1$ and $m_2$, the activities are only weakly correlated as seen by the almost radially symmetric distribution in the $a_1-a_2$ plane. Consequently, the probability distributions for the individual units shown in (b.(ii) - b.(iii)) have largely the same shape as in absence of coupling, see FIG.~\ref{fig:panel1}d. \textit{Simulation parameters: $k = 2000$, $s = 100$, $\omega_{a_{\{1,2\}}} = 50,\,\omega_{m_{\{1,2\}}} = 5, \,T = 0.01,\, m_0 = 4$.}}
    \label{panel5}
\end{figure*}
We now shift the focus from single sensory units to pairs of interacting sensory units as illustrated in FIGs.~\ref{panel5}a.(i) and \ref{panel5}b.(i). The state of a sensory unit with index $i\in\{1,2\}$ is described by the variables $(a_i,m_i)$ that represent the activity and the control (methylation), respectively. The four-dimensional phase space is denoted as $\bar{\Omega} = [0,1]^2\times[0,m_0]^2$. 

\subsection{Overall Adaptation Error}
Before specifying how the units interact, we introduce the adaptation accuracy of a combined system generically. For the coupled pair of systems, the overall activity $a$ and overall error $\epsilon$ are defined as 
\begin{align}
a &\equiv (a_1+a_2)/2,
\label{overallactivity}\\
\epsilon &\equiv \left| 1- \langle a \rangle_{\mathrm{ss}}/a_0 \right|,
\label{OverallError}
\end{align}
where $\langle a \rangle_{\mathrm{ss}}$ represents the stationary-state average of the overall activity and $a_0$ is the setpoint value, which is assumed be the same for both units. By employing the definition of overall activity, Eq.~(\ref{overallactivity}), and the sub-additivity  of the norm, we can establish an upper bound on the overall adaptation error for the interacting system
\begin{equation}
\begin{split}
    \epsilon & = \Bigg| \frac{\langle a_1 \rangle_{\mathrm{ss}} - a_0}{2 a_0} + \frac{\langle a_2 \rangle_{\mathrm{ss}} - a_0}{2 a_0} \Bigg|\\
    & \leq \frac{1}{2}\Bigg| \frac{\langle a_1 \rangle_{\mathrm{ss}} - a_0}{ a_0} \Bigg| + \frac{1}{2}\Bigg|\frac{\langle a_2 \rangle_{\mathrm{ss}} - a_0}{a_0} \Bigg|= \frac{1}{2}(\epsilon_1 + 
             \epsilon_2).
    \label{inequality}
\end{split}
\end{equation}
Here, $\epsilon_1$ and $\epsilon_2$ are the errors of the individual units, which are calculated with respect to the joint probability distribution of the interacting system. The upper bound~(\ref{inequality}) can be generalized to a system comprising an arbitrary finite number of sensory units. 

\subsection{Coupled Activities}
The first interacting system we investigate involves coupling of the activities of two units. Coupling is assumed to be linear in the variables with symmetric coefficients. 
\subsubsection{Model Description}
The system is described by a set of two Langevin equations for each unit, denoted with the indices $i,j\in\{1,2\}$. For unit $i$ and with neighboring unit $j \neq i$ the equations are given by
\begin{equation}
\begin{split}
    \dot{a}_i &= \tilde{F}_{a_i}(a_i,a_j,m_i,s) + \eta_{a_i},\\  
    \dot{m}_i &= F_{m_i}(a_i,m_i,s) + \eta_{m_i},
    \label{CoupledModel1}
\end{split}
\end{equation}
where $\eta_{\{m_i,a_i\}}$ is Gaussian white noise as in Eq.~(\ref{langevinsystem}) and all $\eta_{\ldots}$ with subindices $\{a_1,a_2,m_1,m_2\}$, are statistically independent of each other. The generalized force $\tilde{F}_{a_i}$ that appears in Eqs.~(\ref{CoupledModel1}) is defined as
\begin{equation}
    \tilde{F}_{a_i}(a_i,a_j,m_i,s)  \equiv F_{a_i}(a_i,m_i,s) -  \omega_{ij}(a_i-a_j).
    \label{coupledrive}
\end{equation}
As the interaction between the two units is assumed to be symmetric, we have $\omega_{12}=\omega_{21}$. In Eq.~(\ref{coupledrive}) we set
\begin{equation}
    F_{a_i} = -\omega_{a_i}[a_i - G_i(s,m_i)],
    \label{F_a_i}
\end{equation}
which represents the driving forces that depend on the input signal $s$ through $G_i(s,m_i) = (1+s/e^{2m_i})^{-1}$.\\

The time evolution of the joint probability distribution function $P~\equiv~ P(a_1,m_1,a_2,m_2,t)$ is governed by a Fokker-Planck equation. Conditions for detailed balance in absence of driving can be derived analogously to the non-interacting case, see Supplementary Material. These conditions are consistent with the assumption of one effective temperature for the two units with identical friction and noise strength coefficients. To control the interaction strength, we introduce a tuning parameter $k$ such that
    \begin{equation}
        \omega_{12}  = k(\omega_{a_1}T)= \omega_{21}  = k(\omega_{a_2}T),
    \end{equation}
    where $0\leq k < \infty$ and $\omega_{a_1}=\omega_{a_2}$.
    
Furthermore, the detailed-balance conditions require that 
    $F_{m_i}$ is independent of $a_j$ for $i \neq j$ and $i, j \in \{1,2\}$. 
    Integration of the detailed-balance conditions allows one to systematically construct an expression for $F_{m_i}$ as 
    \begin{equation}
    \begin{split}
        F_{m_i} = \omega_{m_i}(a_i & - a_{0})[\beta \,\,+\\ 
        &(1-\beta)\frac{\Delta_{m_i}}{\Delta_{a_i}}\frac{\omega_{a_i}}{\omega_{m_i}}\partial_{m_i}G_i(s,m_i)]
        \end{split}
        \label{Fm_i}
    \end{equation}
    where $0\leq \beta\leq 1$ and $\frac{\Delta_{m_i}}{\Delta_{a_i}}\frac{\omega_{a_i}}{\omega_{m_i}}=1$ since the effective temperatures are assumed to be equal. We take $\beta = 1$ unless specified otherwise.

The boundary conditions for the system in Eq.~(\ref{CoupledModel1}) are similar to Eq.~(\ref{BC}), but the probability flux is now a four-dimensional vector denoted by $\mathbf{J} = (J_{a_1}, J_{a_2}, J_{m_1}, J_{m_2})^T$. The boundary potential is re-defined accordingly.

The dissipation rate for the interacting system can be obtained from the dissipation rate integral of the system, analogous to Eq.~(\ref{integralDissip}). For system consisting of a pair of sensory units we obtain
\begin{equation}
\begin{split}
   \dot{W} &=  T \sum_{i\in\{1,2\}}\big\langle\big(\Delta_{a_i}^{-1}\tilde{F}_{a_i}\circ\dot{a_i} + \Delta_{m_i}^{-1}F_{m_i}\circ\dot{m_i} \big)\big\rangle_{\mathrm{ss}}.\\
\end{split}
\label{interactingwork}
\end{equation}

\subsubsection{Numerical Results for Coupled Activities}
\begin{figure*}
    \centering
    \includegraphics[width = \linewidth]{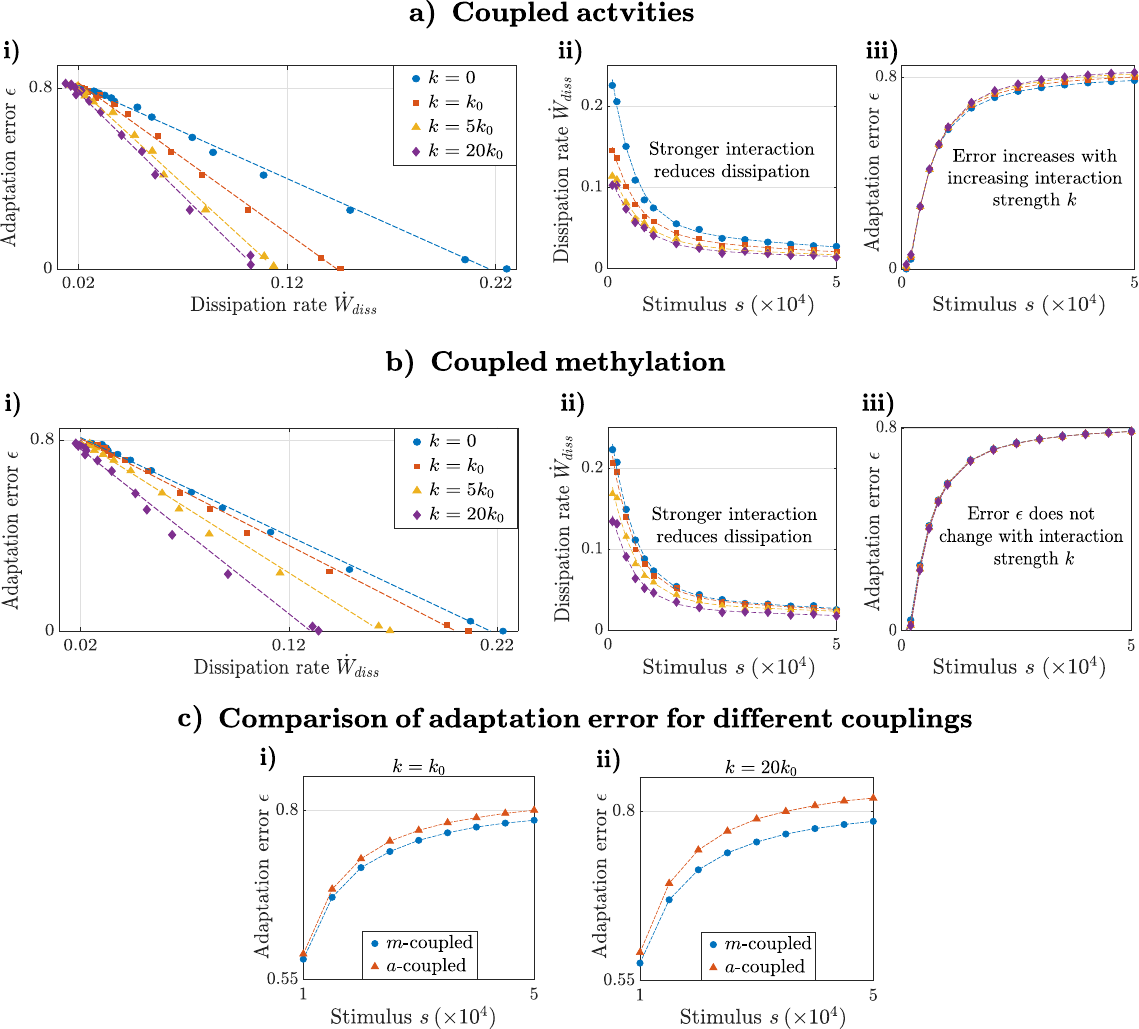}
    \caption{
    Cost-accuracy relation for composite systems consisting of pairs of linearly coupled sensor units.
    \textbf{(a.(i))}~Adaptation error vs. dissipation rate for varying stimulus magnitude $s$ in systems with linearly coupled activities. As for a single, non-interacting unit, the cost-accuracy relation is linear. The parameter $k$ controls the coupling strength. When $k=0$, the units do not interact and the overall dissipation rate is twice the dissipation rate of a single unit. As the coupling strength $k$ increases, less work is required to achieve the same accuracy as without coupling.
  \textbf{(a.(ii), a.(iii))}~Dissipation rate and overall adaptation error as a function of the stimulus magnitude for different coupling strengths $k$. With increasing coupling strengths, the dissipation rate decreases while the adaptation error increases slightly when activities are coupled. \textbf{(b.(i))}~Adaptation error vs. dissipation rate for varying stimulus strength $s$ in systems with linearly coupled control (methylation) variables. For strong coupling, $k \geq 20k_0$, the overall dissipation rate nearly equals that of a single non-interacting system.
  \textbf{(b.(ii), b.(iii))}~Dissipation rate and overall adaptation error as a function of the stimulus magnitude for different coupling strengths $k$. With increasing coupling strength, the dissipation rate decreases while the adaptation error remains unchanged. Linear coupling of the slow control variables thus reduces the energetic cost of adaptation per unit without compromising the adaptive performance.   
  \textbf{(c.(i), c(ii))}~The adaptation error is consistently lower if control variables $m_i$ are coupled, rather than activity variables $a_i$, for otherwise identical systems. Dots are simulation results, dashed lines show data trend. \textit{Simulation parameters: $k_0 = 100$, $\omega_{a_{\{1,2\}}} = 50,\,\omega_{m_{\{1,2\}}} = 5, \,T = 0.01,\, \sigma = 0.01,\, V_0 = 1/2.$}}
    \label{fig:panel6}
\end{figure*}
The set of Langevin equations~(\ref{CoupledModel1}) are solved numerically with the Euler-Maruyama method, as for the non-interacting system. Results are shown in FIGs.~\ref{panel5}a.(ii)-\ref{panel5}a.(v) and FIGs.~\ref{fig:panel6}a. 

The probability distribution function $P$ has four dimensions and plotting the distribution requires a selection of two-dimensional planes in the phase space as shown in FIGs.~\ref{panel5}a.(ii)-\ref{panel5}a.(v). Figures~\ref{panel5}a.(ii) and \ref{panel5}a.(iii) display distributions for two sensory units with strong coupling, i.e., $k\gg 1$. Comparison with results for a single unit, see FIG.~\ref{fig:panel1}d, shows that the coupling distorts the probability distributions since it introduces a correlation between the two activity variables. Consequently, the probability distribution on the $a_1-a_2$ plane aligns along the line $a_1 = a_2$ and the distributions in the two $a-m$ planes also exhibit a stretched variance along the $m-$axis.

Figure~\ref{fig:panel6}a.(i) illustrates the relationship between the overall adaptation error and the dissipation rate for different stimulus magnitudes $s$. As for single sensory units, see FIG.~\ref{fig:panel4}a, the cost-accuracy relation resulting from the numerical solution of Eq.~(\ref{CoupledModel1}) is linear. For vanishing interaction strength, $k=0$, the pair of units together have the same adaptation accuracy as an individual system, albeit with twice the dissipation rate. As $k$ is increased, the slope of the linear relation becomes more negative, implying that interaction reduces the dissipation rate.
Figure~\ref{fig:panel6}a.(iii) illustrates that a linear coupling of the activities reduces accuracy of the system. However, the overall adaptation error increases only slightly while the dissipation is strongly affected by coupling.

\subsection{Coupled Methylation}
After studying coupled activities, we next consider a coupling the control (methylation) variables. Again, the interaction is assumed to be linear and symmetric. 

\subsubsection{Model Description}
In analogy to Eq.~(\ref{CoupledModel1}), the dynamics of a pair of sensory units labeled with the indices $i,j\in\{1,2\}$ and with coupled variables $m_1$~and~$m_2$ is described by Langevin equations. For unit $i$ and with neighboring unit $j \neq i$ the equations are given by
\begin{equation}
\begin{split}
    \dot{a}_i &= F_{a_i}(a_i,m_i,s) + \eta_{a_i},\\  
    \dot{m}_i &= F_{m_i}(a_i,m_i,s) -  \omega_{ij}(m_i-m_j) + \eta_{m_i},
    \label{Coupledmodel2}
\end{split}
\end{equation}
where the white noises $\eta_{\{a_{i},m_{i}\}}$ are again pairwise independent of each other. The generalized forces $F_{a_{i}}$ are the same as in Eq.~(\ref{F_a_i}). To guarantee detailed balance in equilibrium, the system must satisfy six constraints as shown in the Supplementary Material. These constraints are used to construct $F_{m_{i}}$, resulting in the same expression as given in Eq.~(\ref{Fm_i}). The coupling constants are scaled with the effective temperature $T$ as
\begin{equation}
    \omega_{12} = k(\omega_{m_{1}} T) = \omega_{21} = k(\omega_{m_{2}} T),
\end{equation}
where the parameter $0 \leq k<\infty$ is again used to tune the interaction strength.  
Lastly, as the interaction strength is symmetric and linear, the dissipation rate can be again computed using Eq.~(\ref{interactingwork}). 
\subsubsection{Numerical Results for Coupled Methylation}
Results from simulations of the system described by Eq.~(\ref{Coupledmodel2}) are shown in FIGs.~\ref{panel5}b(ii)-\ref{panel5}b.(v) and \ref{fig:panel6}b. The former displays the histogram of trajectories for different variable combinations, representing the probability distribution on different two-dimensional planes. The latter illustrates the cost-accuracy relation of the interacting system for varying interaction strengths.

Typically, the range of the control variable is much larger than the amplitude of the noise, $m_0 \gg \Delta_m$. Therefore, the probability distribution in the $m_1-m_2$ plane appears highly localized, see FIG.~\ref{panel5}b.(v). Coupling the control elements reduces the spread of this distribution on the $m_1-m_2$ plane even more. The two slow variables are strongly correlated. However, the two fast activities are almost independent of each other and their joint distribution is only slightly elongated along the $a_1=a_2$ axis, see FIG.~\ref{panel5}b.(iv). Overall, the probability distribution on the $a_1-m_1$ and $a_2-m_2$ planes are very similar to the distributions for a non-interacting system, FIGs.~(\ref{panel5}b(ii), \ref{panel5}b.(iii)). Therefore, the distributions of the coupled units are distorted after application of a stimulus almost the same way as for a non-interacting system. Adaptation errors of the fully adaptive, interacting system are consequently almost the same as for a single sensory system.

A coupling of the control variables reduces the overall dissipation rate both for trajectories that stay in the phase space interior and for trajectories that interact with the boundaries. For a given coupling strength $k$, the dissipation rate is slightly higher if the $m_i$ are coupled, compared to coupled activities $a_i$.
Figure~\ref{fig:panel6}b.(i) shows a linear relationship between the adaptation error and the dissipation rate required for adaptation, similar to FIG.~\ref{fig:panel6}a.(i). 
Overall, coupling of the control variables reduces the dissipation rate and the adaptation error appears to be independent of interaction strength for this coupling, see FIG.~\ref{fig:panel6}b.(iii).

\subsection{Discussion of Interacting Systems}
To study the adaptive performance for an idealized system consisting of two sensory units, we define an overall adaptation error in terms of the overall activity, see Eqs.~(\ref{OverallError}, \ref{overallactivity}). These definitions result in an upper bound for the overall adaptation error of the system in terms of the individual errors of the interacting sensory units. If one measures the error of individual adaptive units, the overall error must be smaller than or equal to the mean of the individual errors. This inequality is generic as it is robust to the type of interaction between the units and is true even if the systems are not identical, provided they have the same setpoint $a_0$.


The first interacting system that we consider is one in which the activities $a_i$ are coupled, see Eq~(\ref{CoupledModel1}). For this system,
we find that increasing the interaction strength leads to a reduction of the dissipation rate and a slightly worse adaptation accuracy. This change of the system behavior is due to a stretch of the probability distributions for the $(a_i,m_i)$ along the $m_i-$axes. After an increase of the stimulus $s$, the probability distributions shift to higher values of the control variables $m_i$. During this shift, the distributions interact earlier with the boundaries of the phase space if they are stretched. Thus, boundary conditions of the variables influence the behavior of activity-coupled units at smaller stimuli $s$ than for the corresponding non-interacting unit. Instead of coupling the fast activities, we next couple the slow control variables, see Eq.~(\ref{Coupledmodel2}). Similarly to the previous case, the linear, symmetric coupling of the control variables reduces the dissipation rate. However, the state probability distributions for the individual units in the system closely resemble the distribution for a single, uncoupled unit. Therefore, the adaptation accuracy is not compromised by this coupling.

For both types of interacting systems, the finite range of internal variables is irrelevant for small stimulus values $s$ if the states $(a_i,m_i)$ rarely assume their extreme values. In absence of such boundary effects, an increase of the coupling constant reduces the overall dissipation from originally twice the dissipation rate of a single unit down to almost the dissipation rate of only one unit. Thus, a linear interaction of sensory units in the fully adaptive limit reduces the number of wasteful cycles and effectively lowers the energetic cost of adaptation.

\section{Conclusion}
From bacterial chemosensors to yeast osmotic pressure sensors and mammalian light sensors, a wide variety of systems in biology rely on biochemical feedback networks to enable sensory adaptation~\cite{nakatani1991light,muzzey2009systems,Makingsenseofitall}. In the case of \textit{E. coli} chemotaxis, transmembrane receptors sense and adapt to the concentration of extracellular ligands. This process is characterized by a rapid response in sensor activity following a change in ligand concentration and a slowly varying methylation that counteracts the effect of the stimulus and returns the activity to a set point~\cite{tu2008modeling}. 

We employ an idealized model of bacterial adaptation to visualize how the working range affects the accuracy and energetics of adaptation. First, we study the three-node model for adaptation of a single sensory unit. Large stimuli drive the state of the modelled system to its physical limits, represented by boundary conditions of the phase space of internal variables. We find that the adaptation error is a linear function of the dissipation rate. As the strength of a stimulus is increased, the error increases while the dissipation rate decreases. The results from simulations are supported with analytical approximations. Next, we investigate the the same relationship for interacting systems. For pairs of sensory units, the overall adaptation error is determined by the average activity of the two units. An interaction between the units is modelled in two ways. Either by coupling the output activities of the two units or by coupling their control variables representing the methylation state. We find that both types of interactions reduce the dissipation rate and therefore affect the cost-accuracy relationship. Furthermore, the simulation results suggest that a coupling of the methylation variables may be more advantageous than a coupling of the activities because, in this case, the interaction reduces the energetic cost cost of powering multiple units without compromising the accuracy of adaptation. Overall, a cooperative operation of many sensors amplifies sensitivity~\cite{Sourjik_2004} and our models predict that the amplification can be energy efficient since the dissipation rate per sensor is reduced by coupling.

Looking ahead, one can consider to study the energetics of larger systems of coupled units, as well as how they behave in the regime of imperfect adaptation. {Clearly, linear receptor arrays or randomly clustered receptors with next-neighbor couplings may exhibit more complex collective energetics.} From a more formal point of view, it would also be interesting to study how different levels of noise in the different components of the system would affect the adaptive performance while at the same time driving energy flows. Finally, further theoretical and experimental work is needed to assess the biological relevance of a trade-off between the benefit of a large working range and the energetic cost of generating sufficiently complex sensory systems. The analysis of the scenarios involved is far from straightforward, but this work represents a step towards understanding the energetic consequences of a limited working range in adaptation.

\begin{acknowledgments}
This project has received funding from the European Research Council (ERC) under the European Union’s Horizon 2020 research and innovation programme (BacForce, G.A.No. 852585)
\end{acknowledgments}

\nocite{*}

%

\end{document}